\documentclass[]{spie}  
\usepackage[]{graphicx, url, textcomp, amsmath}

\title{On-sky vibration environment for the Gemini Planet Imager and
  mitigation effort}

\author{Markus Hartung\supit{a}, Tom Hayward\supit{a}, Les
  Saddlemyer\supit{b}, Lisa Poyneer\supit{c}, 
Andrew Cardwell\supit{a},
Chas Cavedoni\supit{d},
Myung Cho\supit{e},
Jeffrey~K. Chilcote\supit{f},
Paul Collins\supit{a},
Darren Dillon\supit{g}, 
Ramon Galvez\supit{a}, 
Gaston Gausachs\supit{a},
Stephen Goodsell\supit{d}
Andres Guesalaga\supit{h},
Pascal Hibon\supit{a},
James Larkin\supit{f},
Bruce Macintosh\supit{i},
Dave Palmer\supit{c}, 
Naru Sadakuni\supit{a}, 
Dmitry Savransky\supit{a}, 
Andrew Serio\supit{a},
Fredrik Rantakyro\supit{a}, 
and Kent Wallace\supit{k}
  \skiplinehalf
     \supit{a}Gemini Observatory, La Serena, Casilla 603, Chile;
  \\ \supit{b}National Research Council of Canada Herzberg, Victoria, Canada, United States
  \\ \supit{c}Lawrence Livermore National Laboratory, Livermore, United States
  \\ \supit{d}Gemini Observatory, Hilo, Hawaii, United States
  \\ \supit{e}National Optical Astronomy Observatory, GSMT Program Office, Tucson, United States
  \\ \supit{f}University of California Los Angeles, Los Angeles, United States
  \\ \supit{g}University of California Observatories/Lick Observatory, University of California, Santa Cruz, United States
  \\ \supit{h}Pontificia Universidad Cat\'olica de Chile, Vicuna Mackenna 4860, Santiago, Chile
  \\ \supit{i}Kavli Institute for Particle Astrophysics and Cosmology, Stanford University, Stanford, CA, United States
  \\ \supit{j}Sibley School of Mechanical and Aerospace Engineering, Cornell University, Ithaca, NY 14853, United States
  \\ \supit{k}Jet Propulsion Laboratory/California Institute of Technology, Pasadena, United States
}

\authorinfo{Send correspondence to Markus Hartung: mhartung@gemini.edu, Telephone: +56-51-2205 664}
 
\begin{document} 
  \maketitle 

\begin{abstract}
The Gemini Planet Imager (GPI) entered on-sky commissioning and had
its first-light at the Gemini South (GS) telescope in November
2013. GPI is an extreme adaptive optics (XAO), high-contrast imager
and integral-field spectrograph dedicated to the direct detection of
hot exo-planets down to a Jupiter mass. The performance of the
apodized pupil Lyot coronagraph depends critically upon the residual
wavefront error (design goal of 60\,nm\,RMS with $<$5\,mas RMS
tip/tilt), and therefore is most sensitive to vibration (internal or
external) of Gemini's instrument suite.  Excess vibration can be
mitigated by a variety of methods such as passive or active dampening
at the instrument or telescope structure or Kalman filtering of
specific frequencies with the AO control loop. Understanding the
sources, magnitudes and impact of vibration is key to mitigation. This
paper gives an overview of related investigations based on instrument
data (GPI AO module) as well as external data from accelerometer
sensors placed at different locations on the GS telescope
structure. We report the status of related mitigation efforts, and
present corresponding results.
\end{abstract}



\label{sec:intro}  

\section{GPI's Vibration history} 

   \begin{figure}
   \begin{center}
   \begin{tabular}{c}
   \includegraphics[angle=0,width=0.8\textwidth]{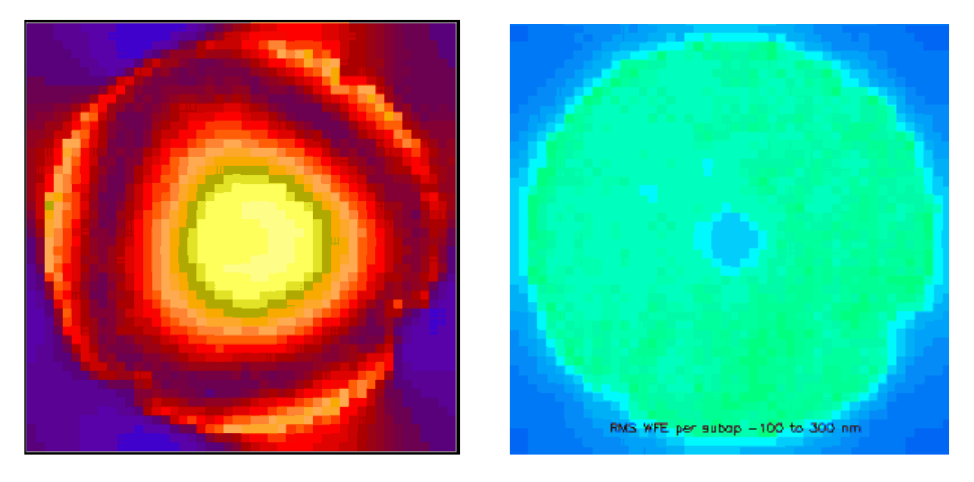}
   \end{tabular}
   \end{center}
   \caption[example] { \label{fig:rms-phase-maps_ccr-on-off}
     Comparison RMS phase maps with cryocoolers on and off. This
     demonstrates that the ``Onigiri''-mode (main component focus) is
     excited in the telescope through the cryocoolers that are mounted
     on GPI's IFS.}
   \end{figure} 

Vibrations have been a concern since early in the GPI
project\cite{macintosh_pnas_2014}. Questions like what are the maximum
allowable vibration levels from the telescope environment or what is
the impact of wind shake (also during target acquisition) had been
specified and discussed between the observatory and GPI collaboration
from early on. Ironically, during on-sky commissioning it turned out
that the most dominant source of vibrations arose from GPI's own
cryocoolers which are injected into the telescope structure. GPI's
Integral Field Spectrograph\cite{larkin_2014} (IFS) is cooled by two
Stirling cycle cryocoolers (Sunpower CryoTel GT). Already in 2011
(when the IFS was still at UCLA) it became clear that the coolers were
a not negligible source of vibrations at 60\,Hz (and harmonics)
causing microphonic effects on the science detector. Design
modifications were undertaken to eliminate microphonics and the
coupling into the IFS was reduced via Sorbothane
washers.\cite{chilcote_2012}. Furthermore, a study was undertaken by
CSA Moog to determine resonance frequencies of GPI\cite{csa_2012}.
During the pre-shipment acceptance test review (July 2013) even though
the injection of vibration from the coolers had been significantly
suppressed, the tip/tilt power spectra showed that we still did not
fully meet the specification ($<$ 4\,mas RMS on a bright star
  excluding measurement noise) but it was demonstrated that this effect
could be mitigated by implementing a Linear-Quadratic-Gaussian (LQG)
controller (Kalman filter)\cite{poyneer_2014}.  The fine tuning of the
LGQ parameters (carefully watching loop stability) continued as soon
as GPI arrived in the Cerro Pachon laboratory (Aug - Oct 2013). Also
during this period we studied carefully the beating effect of the two
cryocoolers. A slight discrepancy (or instability) in the nominal
frequency of 60\,Hz caused the coolers to cycle through an in- and
counter-phase state in a time period of 20 to 30 min producing a
change of the vibration amplitude up to a factor of 4. A changing
gravity vector also seemed to play a role. For these reasons, we
upgraded the Sunpower controller board in January 2014 to a new model
that can drive the two cryocoolers with a constant relative phase.  As
expected, an optimal result is achieved when the coolers are driven
exactly counter phase. With this new board the beating effect
disappeared and the injected vibrations were kept to a minimum.

\begin{table}[b!]
\caption{The mode causing the triangular shape in the RMS phase map
  (dubbed ``Onigiri''-mode). Amplitude and phase relation ship are listed.}
\label{tab:onigiri}
\begin{center}       
\begin{tabular}{|l|l|l|} 
\hline
\rule[-1ex]{0pt}{3.5ex} Zernike Mode & Amplitude (relative to focus) & Phase (rad)  \\
\hline
\rule[-1ex]{0pt}{3.5ex}  Focus     & 1.00 & 0 \\
\hline
\rule[-1ex]{0pt}{3.5ex}  Trefoil (45\textdegree)  & 0.14 & 0 \\
\hline 
\rule[-1ex]{0pt}{3.5ex}  Spherical & 0.15 & $\pi$ \\
\hline 
\end{tabular}
\end{center}
\end{table}

In November and December 2013 we had our first two commissioning runs
at the telescope. During these runs, we focused on basic
functionalities, such as closing the loops (tip/tilt, woofer,
tweeter), loop stability, off-loading to the secondary mirror (M2),
and measurements for the on-sky wavefront error budget. During our
first closed-loop on-sky night in November we were surprised by a
triangular shape in the RMS phase map
(Fig.~\ref{fig:rms-phase-maps_ccr-on-off}). This shape was caused by a
60\,Hz mode in the wavefront that mostly projects into focus. On the
other hand, the 60\,Hz tip/tilt error showed up in predicted levels
and was already successfully mitigated during the March 2014
commissioning run. The tip/tilt LQG filter is meanwhile fully
implemented to be part of regular observations. The filter can be
adjusted with a comprehensive set of parameters (watching loop
stability for an aggressive tuning). Common and non-common path
tip/tilt signals can be addressed as required.\cite{poyneer_2014,
  poyneer_2010}.

The triangular shape in the RMS map was never seen in the laboratory
(using GPI's internal calibration source, the Artificial Star Unit -
ASU). But switching off GPI's cryocoolers removed any doubt that this
signal originates from GPI and is transmitted into Gemini's
instrument support and telescope structure.  During the March 2014
commissioning run we obtained a detailed diagnosis of the 60 Hz
triangular shape on RMS map.  We filtered the AO telemetry at 60\,Hz
to inspect the phase and amplitude relation ship for the lowest
Zernike modes and indeed, we could confirm that the triangular shape
is caused by a mode consisting mainly of focus, trefoil and spherical
(Table~\ref{tab:onigiri}). The mode is dubbed ``Onigiri''-mode because
of its similarity with this Sushi delicacy. The phase shift of $\pi$
for the spherical Zernike mode means a sign flip in respect to the
other modes.
Using the exact same LQG framework as for tip-tilt (but with different
model coefficients) we can correct the 60\,Hz focus signal (the
largest component of the Onigiri-mode). An extra step is needed to
remove the focus signal from the centroids for direct control, see
Poyneer et al.\cite{poyneer_2014} for more detail.\footnote{The CANARY
  team also implemented successfully an LQG controller for higher
  orders.\cite{sivo_2014}} Since the May 2014 run (our last
commissioning run before Gemini South winter shutdown period) the
focus LQG control is implemented and we demonstrated that the 60\,Hz
focus signal can be fully suppressed (below a negligible value of
2\,nm RMS). As a next step we like to experiment with adding the
higher order Zernike terms (trefoil, spherical) and make the
higher-order LQG controller part of standard science operations.

Even though the successful mitigation of LQG filters (tip/tilt and
focus signals) works impressively, there are reasons not to abandon a
``root'' treatment - that is to eliminate the 60\,Hz excitation
directly at GPI's IFS. The high-order wavefront sensor of GPI's
calibration unit\cite{kent_2008} (a Mach-Zehner type interferometer)
cannot be successfully commissioned with these internal vibrations
because of a strong loss in fringe contrast. Furthermore, we are
investigating in how far other instruments (like GeMS) are impacted
and it is always best to keep the telescope vibration environment as
quiet as possible.

  \begin{figure}
   \begin{center}
   \begin{tabular}{c}
   \includegraphics[angle=0,width=1.0\textwidth]{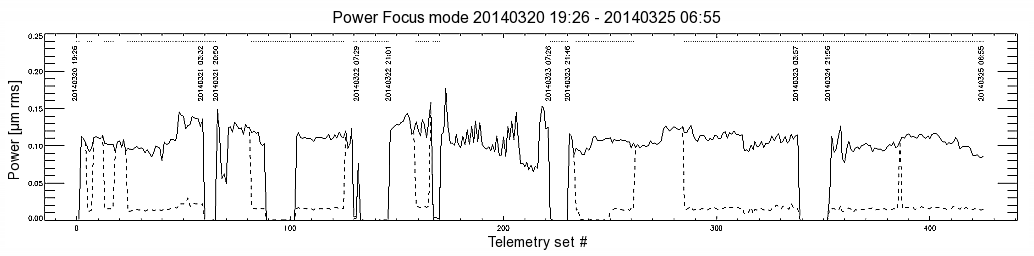}
   \end{tabular}
   \end{center}
   \caption[example] { \label{fig:onigiri-power} The evolution over
     five consecutive commissioning nights (March 2014 run) of the
     power of the 60\,Hz focus signal which is the main constituent of
     the Onigiri-mode.}
   \end{figure}

   \begin{figure}
   \begin{center}
   \begin{tabular}{c}
   \includegraphics[angle=0,width=0.9\textwidth]{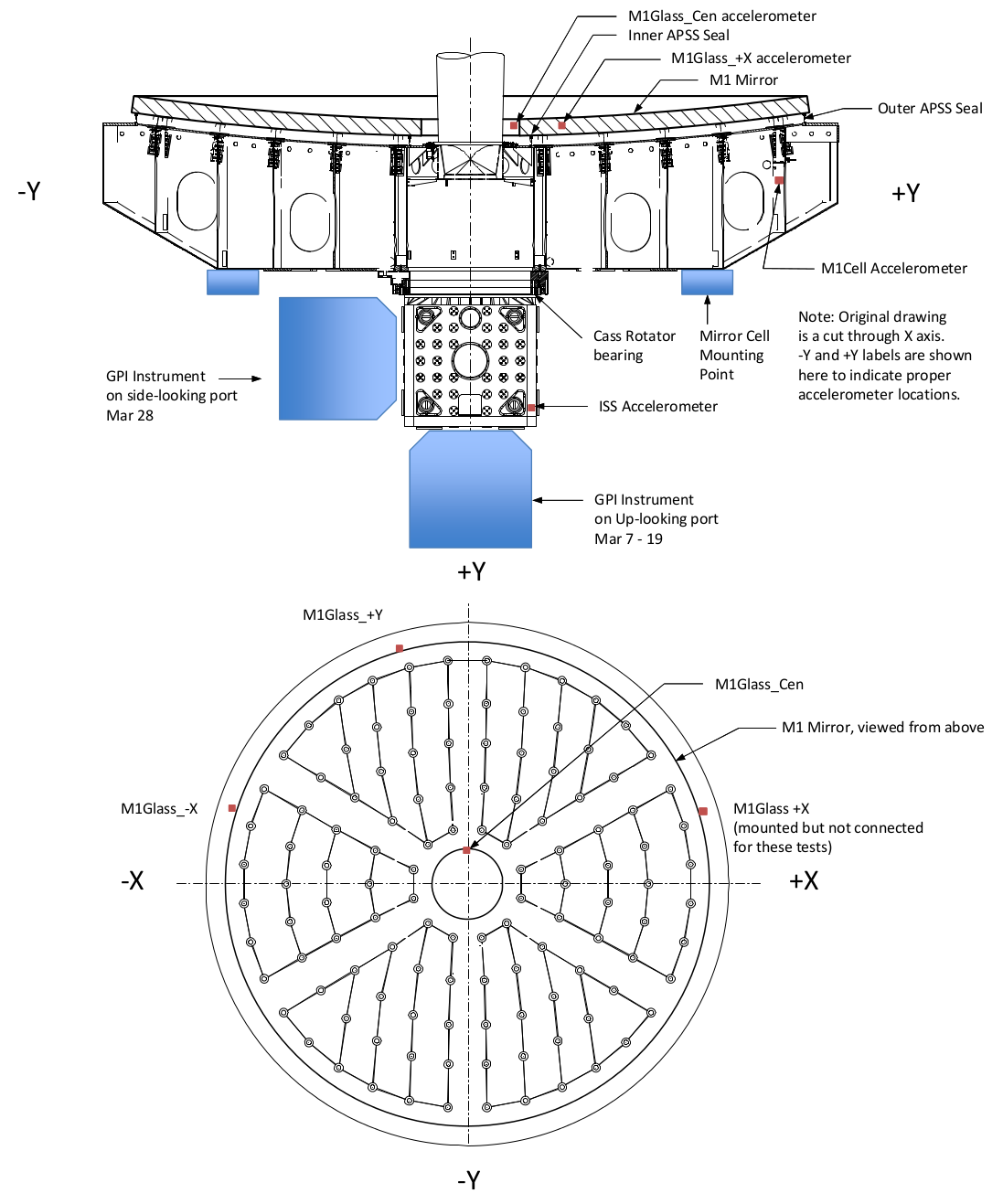}
   \end{tabular}
   \end{center}
   \caption[example] 
   { \label{fig:m1_combined} 
M1 side and top view with locations of accelerometer sensors marked as red squares.}
   \end{figure}

\section{Vibrations: Simulations and measurements}

\subsection{The Onigiri mode} \label{sect:onimode}
In this section we concentrate on the dominant Zernike mode of which
the Onigiri-mode is composed, that is the focus mode
(Table~\ref{tab:onigiri}).  The continuous line in
Fig~\ref{fig:onigiri-power} shows the power of the focus contribution
as it evolved over the March 2014 run.  Therefore, all available AO
telemetry was filtered for 60\,Hz and the focus mode extracted. The
levels range between 80 and 150\,nm RMS. The dates in the plot
indicate the start and the end of the individual commissioning
nights. Intermittently, the signal drops to almost zero - this
corresponds to daytime telemetry on the internal calibration
source. (The dashed line indicates whether the focus mode was included
in the Fourier reconstructor or if the separate focus loop was
active. This information can be ignored for our purpose here.)
Eq.~\ref{eq:df} is used to convert an RMS wavefront error $s$ into a
focus shift $df$ given an f-number of $f/D$:
\begin{equation} \label{eq:df}
df = 16 \sqrt{3} \frac{f}{D} \cdot s
\end{equation}
Thus, a 150\,nm RMS focus wavefront error (WFE) corresponds to a 1\,mm
focus offset at the Cassegrain focus (f/D = 16). The same focus WFE
also arises for a translation of $\pm 13$\,$\mu$m of M2 relative to M1
(or M1 relative to M2, f/D = 1.8).
Nevertheless, be reminded that if no translation were involved to
explain the focus vibration but {\it just} a (parabolic) bending of a
surface in the beam, the corresponding peak to valley sagitta movement
is ``only'' $2\sqrt{3} \cdot 150$\,nm.\footnote{Multiplication by
  $2\sqrt{3}$ converts the RMS value into peak to valley (of a focus
  wavefront in a circular pupil).}  As we describe in the next
section, acceleration measurements show that that the primary mirror (M1) appears to undergo
such a sagitta movement that could explain most of the 60\,Hz-focus vibrations.

\subsection{Accelerometer measurements} 
To investigate the transmission of GPI's 60\,Hz vibration into the
telescope, an extensive set of accelerometer measurements was taken at
different locations of the telescope\cite{hayward_2014}. This study
is on-going and led by the telescope scientist Tom Hayward. It is a
large effort to understand and mitigate vibrations for the Gemini
telescopes in general and is not only GPI related. Here, we present a
small extract of this comprehensive study and some key insights
concerning the transmission of vibrations in the GPI context.

   \begin{figure}
   \begin{center}
   \begin{tabular}{c}
   \includegraphics[angle=0,width=1.0\textwidth]{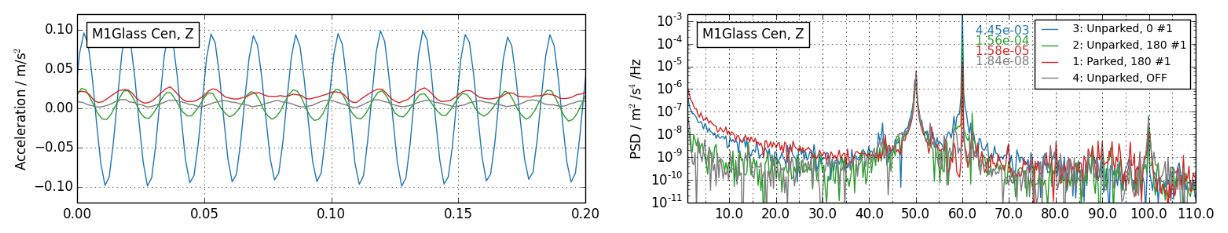}
   \end{tabular}
   \end{center}
   \caption[example] { \label{fig:gpi-M1-vibrations} Measurement
     extracted from GS / GPI vibration study\cite{hayward_2014}. Left:
     Time series, abscissa in sec. Right: Corresponding power spectral
     density, abscissa in Hz. The 60\,Hz signal is dominating.}
   \end{figure} 

A high priority has been to understand and mitigate the excitation of
the Onigiri-mode, particularly its focus contribution, and which
optical surface or surfaces are creating it. GPI's internal surfaces
are excluded since it is not seen internally on the ASU.

First, accelerometer sensors were placed on the M2 support structure. We
measured an acceleration of 0.010\,m/s$^2$ corresponding to a
displacement amplitude of 0.07\,$\mu$m at 60\,Hz.  This is only ~0.5\%
of the required displacement of 13\,$\mu$m (see
Sect.\ref{sect:onimode}) to explain a focus WFE of 150\,nm RMS by a
pure translatory motion. The stiffness of M2 should not allow for a
substantial flexing to create a relevant focus term.

As a next step, we installed accelerometer sensors on the primary
mirror M1. Fig.~\ref{fig:m1_combined} shows a schematic of M1 with the
locations of the sensors at the center and the mirror edge marked as
red squares. The weight of the mirror is supported 120 hydraulic
actuators (marked by circles in the top view) and by an air
cavity. The air cavity sustains 80\% of the mirror's weight and the
actuators only carry the remaining 20\%. The cavity (dubbed ``air
bag'') can be seen in the side view. The inner and outer seals of
the Air Pressure Support System (APSS) are also indicated. In the near
future we will install a dedicate pressure sensor inside the cavity to
investigate in how far vibrations are transmitted via the APSS.

The accelerometer measurements showed that the central part of the
mirror (sagitta) is moving at 60\,Hz against the edges with a measured
maximum acceleration of 0.06\,m/s$^2$ at the mirror center which
corresponds to an amplitude of 0.422\,$\mu$m.  This translates in a
122\,nm RMS wavefront error if we simply interpret this sagitta
movement being solely responsible for the focus wavefront error. But
since we still lack information on the exact phase relation ship
between edge and center sensors this value should be seen as an order
of magnitude value. Nevertheless, this experiment gives striking
evidence that all or the major part of the 60\,Hz focus vibration
originates from a flexing surface of M1.

Fig.~\ref{fig:gpi-M1-vibrations} shows an example of one of the many
time series and power spectra of the Gemini South / GPI vibration
study\cite{hayward_2014}. The dominant line of the power spectra
(right plot) is the 60\,Hz signal from GPI (other lines are at 50 and
100\,Hz). The ratio of the acceleration amplitudes (linear to
Peak-to-Peak displacements) for different test scenarios can be best
compared in the time series on the left side. The largest amplitude
(blue line, legend index 3) is obtained when the two GPI cryocoolers
are tuned to oscillate in phase (0\,deg) and the APSS pressurized to
its default as for regular night time observing (``unparked''). The
green line (legend index 2) corresponds to the two cryocoolers tuned
to counter phase (180\,deg) which minimizes the injection of the
60\,Hz signal. This is also a demonstration of the successful Sunpower
controller upgrade (overseen by L. Saddlemyer and Gausachs) that was
done in February 2014. If the telescope is ``parked'', i.e. if the
``air bag'' is deflated the amplitude goes even further down (red,
legend index 1) and the 60\,Hz contribution disappears completely
(grey line, legend index 4) if the cryocoolers are switched off.

\subsection{Further remarks on the origin of the Onigiri-mode} 
The M1 and M2 acceleration measurements showed that at least the major
part of the focus amplitude originates from M1.  The contributions of
the higher Zernike orders (coma, spherical) are small (approx. 20\,nm
RMS or smaller) and the triangular shape is still not fully
understood.

First, we suspected M2 to be the source of the triangular shape since
it is mounted with three links (120 deg apart) to the M2 support
structure. On the other hand, as seen on the top view of M1
(Fig.~\ref{fig:m1_combined}), the hydraulic actuators are organized
into six segments which are pairwise interconnected which might also
explain the triangular geometry.  Further acceleration measurements at
additional locations on M1 and M2 would be needed to completely
understand the origin of the triangular shape. In practice these
measurements require significant resources. This underlines the
importance to have a suite of (ideally triaxial) acceleration sensors
permanently mounted at the telescope with sufficient read-out
channels.

   \begin{figure}[t!]
   \begin{center}
   \begin{tabular}{c}
   \includegraphics[angle=0,width=0.4\textwidth]{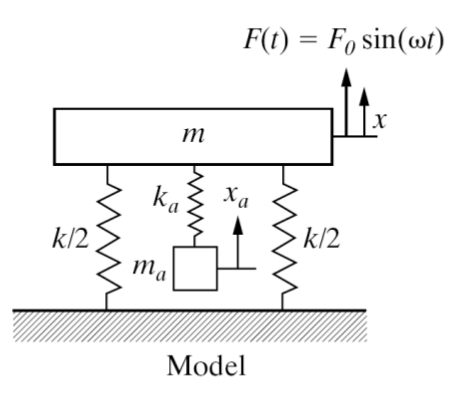}
   \end{tabular}
   \end{center}
   \caption[example] 
   { \label{fig:model-vib-sketch} 
Compensation of an harmonic external force onto a primary mass $m$ using an absorber mass $m_a$.}
   \end{figure}

\subsection{FEA study of primary mirror response to an APSS force} 
In the context of GPI's 60\,Hz vibrations, a Finite Element Analysis
(FEA) was done by Myung Cho to study the effect of the APSS on the
mirror figure.\cite{cho_2014} Here we give a short summary of
preliminary key results of this effort which is also substantial to
understand Gemini's vibration environment in general and to develop
suitable mitigation strategies. The model aims to include all hardware
mounted on the outer edge of the mirror, the air cavity seals at their
nominal positions and with their nominal forces (outer seal 283\,N/m,
inner seal 484\,N/m). Furthermore, axial support locations, forces,
and the effect of the air pressure in the cavity ("air bag") is
included. These are the preliminary results based on the FEA model:

In the static case, to obtain a WFE of 100 nm RMS requires a force per
meter of 1.6\,N/m or 8.1\,N/m on the outer or inner seal,
respectively. This corresponds to a total force of 40.5\,N or 37.3\,N
on the the outer or inner seal, respectively.

The dynamic amplification yields approximately a factor of 7 and the
required excitation force to generate a 100 nm RMS WFE is reduced to
approximately 5.5\,N for the inner {\it or} outer seal. (The WFE on
the mirror figure is much smaller if inner and outer seal are excited
in phase with the same force.)

To conclude, this FEA study shows that a dynamic force (60\,Hz) as
small as 5.5\,N on the APSS is sufficient to cause the seen WFE. It is
plausible that the excitation forces of GPI's cryocoolers can easily
account for this transmitted force and this study is also consistent
with the accelerometer results.

\section{A short theory reminder vibration absorbers and terminology} 
\label{sect:theory}

   \begin{figure}
   \begin{center}
   \begin{tabular}{c}
   \includegraphics[angle=0,width=0.6\textwidth]{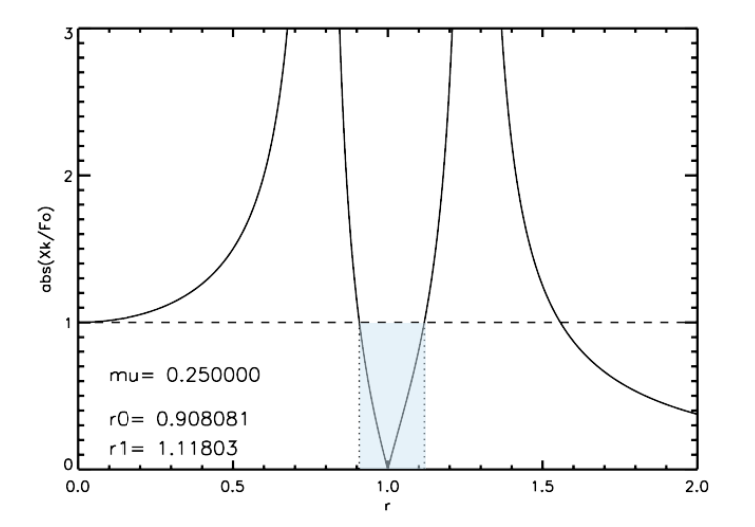}
   \end{tabular}
   \end{center}
   \caption[example] { \label{fig:mu_0p25-shaded-square} The
     normalized displacement of the primary mass vs normalized
     frequency. The shaded square indicates the frequency range in
     which the absorber system actually reduces the
     displacement. Outside of these frequencies amplification occurs.}
   \end{figure}

In the concept of an harmonic oscillator the natural angular frequency
is given by $\omega_p = \sqrt{k/m}$ ($p$ for primary system) where $k$
is the spring constant and $m$ the mass.\footnote{This is common
  lecture book knowledge, e.g. see the book of Hartog on mechanical
  vibrations\cite{hartog_1985}. The Internet provides plenty of
  material, such as free MIT online lectures on
  vibrations\cite{mit_2014} or the website or Russel, Penn. State
  University\cite{russel_2014}.} This is the only frequency a 1-degree
of freedom system (dof) oscillates at if no other forces are applied. If
an oscillating external force acts on a primary mass as shown in
Fig.~\ref{fig:model-vib-sketch} then this force can be compensated for
by attaching an absorber mass to the primary mass. This converts a
1-dof system into a 2-dof system with two eigen-frequencies slightly
different from the natural frequencies of the separate systems. If the
natural frequency of the absorber mass $\omega_a = \sqrt{k_a/m_a}$
($a$ for absorber system) exactly matches the external force frequency
$\omega$, the forces cancel out and the displacement amplitude of the
primary mass is zero.

Substituting a synchronous motion (Eq.~\ref{eq:sync}) into the
equations of motion (Eq.~\ref{eq:eom})
\begin{equation} \label{eq:sync}
	\left.\begin{aligned}
		x(t) = X \sin \omega t \quad\quad \\
		x_a(t) = X_a \sin \omega t \quad
	\end{aligned}
	\right\}
	\text{Sub. into EOM}
\end{equation}
\begin{equation} \label{eq:eom}
	\begin{bmatrix}
		m & 0 \\
		0 & m_a
	\end{bmatrix}
	\begin{Bmatrix}
		\ddot{x} \\
		\ddot{x}_a
	\end{Bmatrix}
	+
	\begin{bmatrix}
		k+k_a & -k_a \\
		-k_a & k_a
	\end{bmatrix}
	\begin{Bmatrix}
		x \\
		x_a
	\end{Bmatrix}
	=
	\begin{Bmatrix}
		F_0 \sin \omega t \\
		0
	\end{Bmatrix}
\end{equation}
yields a displacement amplitude $X$ of
\begin{equation}
X = \frac{ (k_a - m_a \omega^2) F_0 }{ (k + k_a - m \omega^2)(k_a-m_a \omega^2) - {k_a}^2 }
\end{equation}
or a normalized displacement of the primary mass of
\begin{equation} \label{eq:nomarlized-displacement}
\left| \frac{Xk}{F_0} \right| = \left| \frac{ 1-r^2 }{ (1 + \mu \beta^2 - r^2)(1-r^2) - \mu \beta^2} \right|
\end{equation}
where $\mu = m_a/m$, $\beta = \omega_a / \omega_p$ and $r = \omega /
\omega_a$. The normalized displacement
(Eq.~\ref{eq:nomarlized-displacement}) is shown in
Fig.~\ref{fig:mu_0p25-shaded-square}. The shaded square indicates the
useful operating bandwidth of the absorber system ($0.91 \omega_a <
\omega < 1.12 \omega_a$) for a specific absorber mass to mass ratio of
1/4 ($\mu = 0.25)$. Recommended mass ratios range between $0.05 < \mu <
0.25.$ The higher the mass ratio (i.e. the larger the absorber mass)
the larger the usable bandwidth but for practical reasons and
engineering constraints the absorber mass is usually kept as small as
possible. Table~\ref{tab:mu} lists ``useful'' bandwidths for a few
different mass ratios. Particularly for small absorber masses (small
$\mu$) it is important to remind that {\it the absorber system can
  amplify vibrations if not carefully tuned.}

\begin{table}
\caption{Normalized minimum and maximum frequencies $r$ and useful bandwidth for different mass ratios. To obtain regular (unnormalized) frequencies multiply $r_{min}$ and $r_{max}$ by $\omega_a$ (natural frequency of absorber system.} 
\label{tab:mu}
\begin{center}       
\begin{tabular}{|c|c|c|c|} 
\hline
\rule[-1ex]{0pt}{3.5ex} mass ratio $\mu$ & $r_{min}$ & $r_{max}$ & useful bandwidth (\%) \\
\hline
\rule[-1ex]{0pt}{3.5ex} 0.01 & 0.995 & 1.005  & 0.99 \\
\hline
\rule[-1ex]{0pt}{3.5ex} 0.05 & 0.977 & 1.025  & 4.78 \\
\hline                       
\rule[-1ex]{0pt}{3.5ex} 0.10 & 0.957 & 1.049  & 9.19 \\
\hline                       
\rule[-1ex]{0pt}{3.5ex} 0.25 & 0.908 & 1.118  & 21.0 \\
\hline 
\end{tabular}
\end{center}
\end{table}

\section{Vibration Mitigation with Hardware}
\subsection{Tunable Vibration Absorbers \& Tuned Mass Dampers}

   \begin{figure}[b!]
   \begin{center}
   \begin{tabular}{c}
   \includegraphics[angle=0,width=0.6\textwidth]{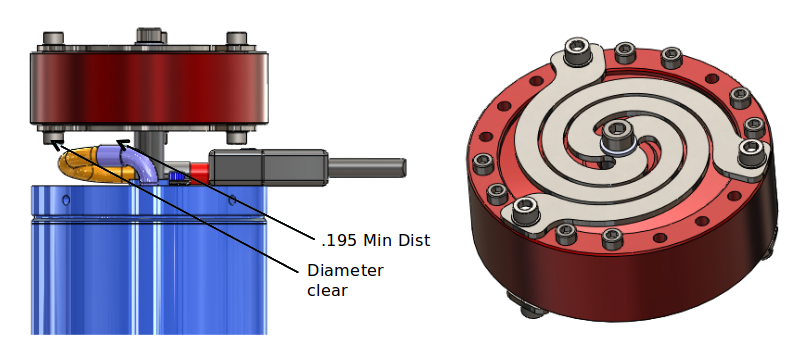}
   \end{tabular}
   \end{center}
   \caption[example] 
   { \label{fig:tva_combined} 
A TVA from CSA Moog as mounted on one of GPI's cryocoolers (left) and a top view of the TVA only (right).}
   \end{figure}

In the short theory section we reminded about the basics to mitigate
vibrations by attaching an {\it undamped} absorber mass system. Such
systems are typically referred to as Tunable Vibration Absorbers
(TVA). They are a good choice if the vibration response is dominated
by a single frequency (nearby motor, pump or other disturbance
source). This frequency should be separated far enough from a
structural resonance frequency otherwise a Tuned Mass Damper (TMD) is
likely a better choice.

TMDs are suited to reduce modal frequencies of a structure that are
excited by any means (motors, mechanical shock, wind, etc). The
difference between a TMD and a TVA is that a damping term $b \dot{x}$
is included:
\begin{equation}
m \ddot{x} + b \dot{x} + k x = F(t),
\end{equation}
whereby F(t) is the external excitation force and $b \dot{x}$ the term
describing the dampening. This term is physically realized by an
element that dissipates energy. It could be simply a visco-elastic
medium or a using a magnetic device similar to a voice coil. The
second approach has the advantage that it can also be used for an
active compensation.

   \begin{figure}
   \begin{center}
   \begin{tabular}{c}
   \includegraphics[angle=0,width=1.0\textwidth]{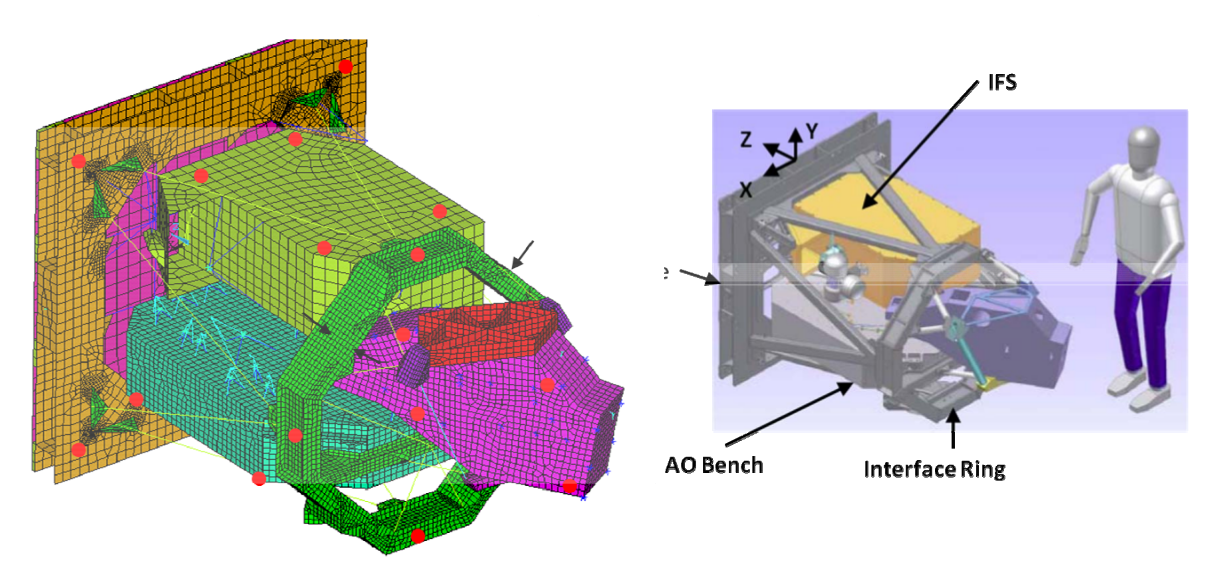}
   \end{tabular}
   \end{center}
   \caption[example] { \label{fig:csa-study-cominbed} GPI wire frame model
     (left) and coordinate system with excitation (black arrows) and
     measurements points (red dots) for modal frequency study
     (right).}
   \end{figure}

\begin{table}
\caption{Specifications from CSA Moog for the TVA for GPI's cryocoolers}
\label{tab:tva}
\begin{center}       
\begin{tabular}{|l|l|} 
\hline
\rule[-1ex]{0pt}{3.5ex} Frequency & 60\,Hz (tuning screws 59.5 to 60.5\,Hz)  \\
\hline
\rule[-1ex]{0pt}{3.5ex} Moving Mass &  2.26 lbs (1.03 kg) \\
\hline
\rule[-1ex]{0pt}{3.5ex} Total Mass & 2.26 lbs (1.03 kg)  \\
\hline 
\rule[-1ex]{0pt}{3.5ex} Max Stroke (hard limit) & $\pm$ 0.135'' (3.43 mm) \\
\hline 
\rule[-1ex]{0pt}{3.5ex} Estimated operational stroke & $<$ 0.100'' (2.54 mm) \\
\hline 
\rule[-1ex]{0pt}{3.5ex} Force reacted at 60Hz and $\pm$ 0.100'' &  334 N \\
\hline 
\rule[-1ex]{0pt}{3.5ex} Force reacted at 60Hz and $\pm$ 0.135'' &  467 N \\
\hline
\end{tabular}
\end{center}
\end{table}

Fig.~\ref{fig:tva_combined} shows one of the two GPI cryocoolers with
a TVA from CSA Moog mounted. Its specifications are listed in
Tab.~\ref{tab:tva}. The TVA is specified for 60\,Hz and can be
finetuned by $\pm$\,0.5\,Hz using the depicted tuning screw. As
stated in Sect.~\ref{sect:theory} the fine tuning is critical for
absorbing efficiency. In the worst case amplification can occur,
particularly if a natural structural mode is located in the close
neighborhood.  It is notable (Table~\ref{tab:tva}) that even though
the TVA mass is only about one kilogram and has a maximum stroke of a three
and a half mm, a reacting force\footnote{The reacting force corresponds
  to the centripetal force with an angular frequency $\omega$ and a
  ``stroke'' $r$: $F = m \omega^2 r$.} up to 467\,N is created at
60\,Hz (corresponding to 48\,kg if translated into weight). After
shipment and arrival of GPI to Cerro Pachon in Chile, it appeared that
the TVAs lost their effectiveness which is currently investigated by
CSA (and the TVAs were dismounted from the cryocooler heads for
inspection). For the next commissioning run in September 2014 we
expect these TVAs to be mounted again with a corrected fine-tuning.

\subsection{GPI modal study}

\begin{table}[b!]
\caption{GPI stand-alone modes with suggested TMD specifications.}
\label{tab:gpi-modes}
\begin{center}       
\begin{tabular}{|c|l|r|r|r|r|l|} 
\hline
\rule[-1ex]{0pt}{3.5ex}  Mode & \multicolumn{1}{|c|}{TMD}  & \multicolumn{1}{|c|}{Modal} & \multicolumn{1}{|c|}{f (Hz)} & \multicolumn{1}{|c|}{TMD moving} & \multicolumn{1}{|c|}{$\mu$} & \multicolumn{1}{|c|}{Comment} \\
\rule[-1ex]{0pt}{3.5ex}   & \multicolumn{1}{|c|}{Location} & \multicolumn{1}{|c|}{mass (kg)} &  & \multicolumn{1}{|c|}{mass (kg)} &  &  \\
\hline
\rule[-1ex]{0pt}{3.5ex} 1 & Interface Ring & 2,276 & 48.7 & 22.8 & 0.01 & GPI frame torsion about Y \\
\hline
\rule[-1ex]{0pt}{3.5ex} 2 & Interface Rings & 2,172 & 52.2 & 21.7 & 0.01 & GPI frame rocking about Y \\
\hline 
\rule[-1ex]{0pt}{3.5ex} 3 & CAL & 213 & 74.6 & 2.1 & 0.01 & CAL plunging in Z against rocking IFS \\
\hline 
\rule[-1ex]{0pt}{3.5ex} 4 & CAL & 450 & 80.4 & 4.5 & 0.01 & IFS bouncing against GPI frame \\
\hline 
\rule[-1ex]{0pt}{3.5ex} 5 & CAL & 690 & 93.8 & 6.9 & 0.01 & CAL rocking about Y \\
\hline 
\rule[-1ex]{0pt}{3.5ex} 6 & CAL & ? & 114.9 & ? & & CAL torsion \\
\hline
\rule[-1ex]{0pt}{3.5ex} 7 & CAL & 422 & 126.5 & 4.2 & 0.01 & CAL and IFS rocking about X, Z \\
\hline
\end{tabular}
\end{center}
\end{table}

CSA Moog was contracted to study GPI's modal
frequencies\cite{csa_2012}. They prepared a wire frame model for
simulation. Test measurements were done using a modal hammer to excite
the GPI structure and accelerometer sensors were attached to measure
the response.  Fig.~\ref{fig:csa-study-cominbed} visualizes the wire
frame model. The excitation locations are marked with black arrows;
one arrow is hidden behind the purple CAL unit. The red dots indicate
the locations of the accelerometer sensors. Tab.~\ref{tab:gpi-modes}
shows the extracted modes from their data set in the frequency range
from 40 to 150\,Hz. All recommended masses for the TMDs by CSA are
based on a mass ratio of $\mu=0.01$.  The study was done in December
2012 while GPI was in the A\&T laboratory\cite{hartung_2013} at UCSC
and mounted to a carriage creeper (dubbed the ``L-frame''). The first
two modes with the lowest frequencies (49 and 52\,Hz) are likely to
shift when GPI is attached to the telescope. This is one of the
reasons why no TMDs have been installed yet. The decision was taken to
revisit this study once enough on-telescope experience is gained to
evaluate if TMDs are required. To this point no TMDs are installed and
so far it seems that no TMDs are required. The main vibration line in
our error budget (60\,Hz) is not on a structural mode assuming that
the two low frequency structural modes of GPI have not shifted through
the mounting at the telescope in an unfortunate way.

\section{Summary}
The topic of vibration in general is complex and is a topic prone to
misunderstandings between mechanics, opticians, electrical engineers
and scientists. An awareness of context and language is critical in
this interdisciplinary field. Coordinating a sensible and successful
vibration mitigation effort in an effective way is a challenge and
requires its resources and roles such as an interconnected project
manager and a ``vibration officer''. There is a constant struggle with
hard decisions: Does it make sense to spend the extra dollar to finish
a study that runs out of funding? And avoiding the ``sunk cost'' effect
in a ``prototype'' development environment is never easy.

GPI has the strongest constraints on its wavefront error budget of all
Gemini South instruments and certainly has risen the observatory's
awareness of vibrations in the telescope environment. The mitigation
using an LQG controller for both pointing and focus has been
implemented successfully. This is an intriguing demonstration of how
new technologies can be used to mitigate vibration effects from the
controller side. In that sense, using the LQG controller, the impact
of the cryocooler vibrations on GPI's performance is low if not
negligible.  The only exception is the high-order wavefront front
sensor of the CAL unit. It is a Mach-Zehnder type interferometer and
the loss in fringe contrast due to the vibration prevents its
commissioning.  Fortunately, this is not critical, since the basic
functionalities of the CAL (keeping the star bore-sight, and low-order
wavefront sensing) are not impacted. Furthermore, the GPI team has
developed further alternatives to treat NCPA such as
speckle-nulling. And so far, even though other instruments such as
GeMS {\it can} see the vibration signal on their wavefront sensors, it
seems that there is no impact on performance\footnote{We are still
  waiting for a confirmation of this statement for excellent seeing
  conditions. For medium or worse seeing conditions, we confirmed
  already that performance is not impacted}.  Nevertheless, a ``root
treatment'' is always preferable and it is in the observatory's
natural interest to keep its ``vibration'' spectra as clean as
possible. Recently, Sunpower has developed an Active Cancellation
System (ACS) as add-on for its GT cryocooler series and we are
investigating to upgrade our coolers with it.

There is widespread use of TMDs for aircrafts, machinery, in the car
industry, for trains and railways, buildings, power lines and in
optical systems. And certainly there is still plenty of room to make a
better use of these devices in the field of astronomical
instrumentation and optical telescopes.

A good infrastructure of acceleration sensors is required to
effectively understand and solve unwanted vibrations. Otherwise
measurements are difficult to obtain and to confirm, and resources of
the observatory are wasted in large overheads. In the case of the
Gemini Observatory it seems recommendable to have at least 5 sensors
around the cell of M1 (and one could dream about having each hydraulic
M1 actuator equipped with a sensor). ELTs will take this matter
(hopefully) serious, plan for an integrated design and maybe feature a
dedicated ``vibration control room''.

As GPI and other (XAO) instruments can always be seen as pathfinders
for future telescopes and instrumentation, sharing our challenges is
hopefully helpful to find suitable solutions in other complex
environments.  GPI can serve as a good example for the successful
implementation of a mitigation strategy and for how powerful
mitigation strategies such as LGQ filtering of the wavefront have
become.

\acknowledgments 
The authors thank Paul Langlois for suggestions and improvements on
the figures.  The Gemini Observatory is operated by the Association of
Universities for Research in Astronomy, Inc., under a cooperative
agreement with the NSF on behalf of the Gemini partnership: the
National Science Foundation (United States), the National Research
Council (Canada), CONICYT (Chile), the Australian Research Council
(Australia), Minist\'{e}rio da Ci\^{e}ncia, Tecnologia e
Inova\c{c}\~{a}o (Brazil) and Ministerio de Ciencia, Tecnolog\'{i}a e
Innovaci\'{o}n Productiva (Argentina).



\begin{thebibliography}{10}

\bibitem{macintosh_pnas_2014}
{Macintosh}, B., {Graham}, J.~R., {Ingraham}, P., {Konopacky}, Q., {Marois},
  C., {Perrin}, M., {Poyneer}, L., {Bauman}, B., {Barman}, T., {Burrows}, A.,
  {Cardwell}, A., {Chilcote}, J., {De Rosa}, R.~J., {Dillon}, D., {Doyon}, R.,
  {Dunn}, J., {Erikson}, D., {Fitzgerald}, M., {Gavel}, D., {Goodsell}, S.,
  {Hartung}, M., {Hibon}, P., {Kalas}, P.~G., {Larkin}, J., {Maire}, J.,
  {Marchis}, F., {Marley}, M., {McBride}, J., {Millar-Blanchaer}, M.,
  {Morzinski}, K., {Norton}, A., {Oppenheimer}, B.~R., {Palmer}, D.,
  {Patience}, J., {Pueyo}, L., {Rantakyro}, F., {Sadakuni}, N., {Saddlemyer},
  L., {Savransky}, D., {Serio}, A., {Soummer}, R., {Sivaramakrishnan}, A.,
  {Song}, I., {Thomas}, S., {Wallace}, J.~K., {Wiktorowicz}, S., and {Wolff},
  S., ``{First Light of the Gemini Planet Imager},'' {\em Proceedings of the
  National Academy of Sciences}  (2014).

\bibitem{larkin_2014}
{Larkin}, J.~E., {Chilcote}, J.~K., {Aliado}, T., {Bauman}, B.~J., {Brims}, G.,
  {Canfield}, J.~M., {Cardwell}, A., {Dillon}, D., {Doyon}, R., {Dunn}, J.,
  {Fitzgerald}, M.~P., {Graham}, J.~R., {Goodsell}, S., {Hartung}, M., {Hibon},
  P., {Ingraham}, P., {Johnson}, C.~A., {Kress}, E., {Konopacky}, Q.~M.,
  {Macintosh}, B.~A., {Magnone}, K.~G., {Maire}, J., {McLean}, I.~S., {Palmer},
  D., {Perrin}, M.~D., {Quiroz}, C., {Rantakyr{\"o}}, F., {Sadakuni}, N.,
  {Saddlemyer}, L., {Serio}, A., {Thibault}, S., {Thomas}, S.~J., {Vallee}, P.,
  and {Weiss}, J.~L., ``{The Integral Field Spectrograph for the Gemini Planet
  Imager},'' in {\em Ground-based and Airborne Instrumentation for Astronomy
  V}{\nolinebreak\hspace{0.1em}},  {\em Proc. SPIE} {\bf 9147} (2014).

\bibitem{chilcote_2012}
{Chilcote}, J.~K., {Larkin}, J.~E., {Maire}, J., {Perrin}, M.~D., {Fitzgerald},
  M.~P., {Doyon}, R., {Thibault}, S., {Bauman}, B., {Macintosh}, B.~A.,
  {Graham}, J.~R., and {Saddlemyer}, L., ``{Performance of the integral field
  spectrograph for the Gemini Planet Imager},'' in {\em Ground-based and
  Airborne Instrumentation for Astronomy IV}{\nolinebreak\hspace{0.1em}},
  {\em Proc. SPIE} {\bf 8446} (Sept. 2012).

\bibitem{csa_2012}
Pargett, T., ``{Gemini Planet Imager (GPI) Modal Test Report},'' CSA Report No.
  2012712 (Dec. 2012).

\bibitem{poyneer_2014}
Poyneer, L.~A., {De Rosa}, R.~J., Macintosh, B., Palmer, D.~W., Perrin, M.~D.,
  Sadakuni, N., Savransky, D., Bauman, B., Cardwell, A., Chilcote, J.~K.,
  Dillon, D., Gavel, D., Goodsell, S.~J., Hartung, M., Hibon, P., Rantakyr\"o,
  F.~T., Thomas, S., and V\'{e}ran, J.-P., ``On-sky performance during
  verification and commissioning of the {G}emini {P}lanet {I}mager's adaptive
  optics system,'' in {\em Adaptive Optics Systems
  IV}{\nolinebreak\hspace{0.1em}},  {\em Proc. SPIE {\bf 9148}} (2014).

\bibitem{poyneer_2010}
Poyneer, L.~A. and V\'{e}ran, J.-P., ``Kalman filtering to suppress spurious
  signals in adaptive optics control,'' {\em J. Opt. Soc. Am. A}~{\bf 27}(11),
  A223--A234 (2010).

\bibitem{sivo_2014}
Sivo, G., Kulcs\'ar, C., Conan, J.-M., Raynaud, H.-F., Gendron, E., Basden, A.,
  Vidal, F., Morris, T., Meimon, S., Petit, C., Gratadour, D., Martin, O.,
  Hubert, Z., Sevin, A., Perret, D., Chemla, F., Rousset, G., Dipper, N.,
  Talbot, G., Younger, E., Myers, R., Henry, D., Todd, S., Atkinson, D.,
  Dickson, C., and Langmore, A., ``First on-sky {SCAO} validation of full {LQG}
  control with vibration mitigation on the {CANARY} pathfinder,'' {\em Optics
  Express}  (2014).

\bibitem{kent_2008}
{Wallace}, J.~K., {Angione}, J., {Bartos}, R., {Best}, P., {Burruss}, R.,
  {Fregoso}, F., {Levine}, B.~M., {Nemati}, B., {Shao}, M., and {Shelton}, C.,
  ``{Post-coronagraph wavefront sensor for Gemini Planet Imager},'' in {\em
  Adaptive Optics Systems}{\nolinebreak\hspace{0.1em}},  {\em Proc. SPIE} {\bf
  7015} (July 2008).

\bibitem{hayward_2014}
Hayward, T., ``{Gemini South / GPI Vibration Test Report},'' Rev. 3, Gemini
  Observatory, internal report (May 2014).

\bibitem{cho_2014}
Cho, M., ``{Gemini Primary Mirror Response to GPI vibration - M1 airbag effects
  - Static and dynamic responses},'' internal report (2014).

\bibitem{hartog_1985}
Hartog, D.,  {\em Mechanical Vibrations}{\nolinebreak\hspace{0.1em}}, Dover
  (1985).

\bibitem{mit_2014}
{Massachusetts Institute of Technology}, ``{MIT OpenCourseWare}.''
  \url{http://ocw.mit.edu/courses/find-by-topic/}.

\bibitem{russel_2014}
Russel, ``Acoustics and vibration animations.''
  \url{http://www.acs.psu.edu/drussell/}.

\bibitem{hartung_2013}
{Hartung}, M., {Macintosh}, B., {Poyneer}, L., {Savransky}, D., {Gavel}, D.,
  {Palmer}, D., {Thomas}, S., {Dillon}, D., {Chilcote}, J., {Ingraham}, P.,
  {Sadakuni}, N., {Wallace}, K., {Perrin}, M., {Marois}, C., {Maire}, J.,
  {Rantakyro}, F., {Hibon}, P., {Saddlemyer}, L., and {Goodsell}, S., ``{Final
  A\&T stages of the Gemini Planet Finder},'' in {\em {Proceedings of the
  Third AO4ELT Conference}}{\nolinebreak\hspace{0.1em}},  {Esposito}, S. and
  {Fini}, L., eds. (Dec. 2013).

\end{thebibliography}

\bibliographystyle{spiebib}   

\end{document}